\documentclass[runningheads]{llncs}
\usepackage[T1]{fontenc}
\usepackage{graphicx}
\usepackage{amsmath,amssymb}
\usepackage{tikz}
\usepackage[table]{xcolor}
\usepackage{subcaption}
\usepackage{booktabs}
\usepackage{multirow}
\usepackage[most]{tcolorbox}
\usepackage{placeins}
\usepackage{flafter}
\usepackage{float}
\usepackage{makecell}
\usepackage[ruled,vlined,linesnumbered]{algorithm2e}

\usepackage[
    colorlinks=true,
    linkcolor=blue,
    citecolor=blue,
    urlcolor=blue
]{hyperref}

\newtcolorbox{rqbox}{
    colback=gray!15,
    colframe=black,
    boxrule=0.5pt,
    arc=0pt,
    left=4pt,
    right=4pt,
    top=4pt,
    bottom=4pt,
    before skip=6pt,
    after skip=6pt
}

\newcommand{\circnum}[1]{\tikz[baseline=(char.base)]{
\node[shape=circle,draw,inner sep=1pt] (char) {\small #1};}}
\begin{document}
\title{RAMP: Reversing Adversarial Perturbations to Strengthen Clean-Label Backdoor Attacks against Malware Detectors}
\titlerunning{RAMP against Malware Detectors}
% If the paper title is too long for the running head, you can set
% an abbreviated paper title here
%
% \author{First Author\inst{1}\orcidID{0000-1111-2222-3333} \and
% Second Author\inst{2,3}\orcidID{1111-2222-3333-4444} \and
% Third Author\inst{3}\orcidID{2222--3333-4444-5555}}
\author{Jinwen Xin \and
Dongni Zhang \and
Chenyang Wang \and
Jianming Fu \and
Ming Tang \and
Guojun Peng}
\authorrunning{J. Xin et al.}
% First names are abbreviated in the running head.
% If there are more than two authors, 'et al.' is used.
%
% \institute{Princeton University, Princeton NJ 08544, USA \and
% Springer Heidelberg, Tiergartenstr. 17, 69121 Heidelberg, Germany
% \email{lncs@springer.com}\\
% \url{http://www.springer.com/gp/computer-science/lncs} \and
% ABC Institute, Rupert-Karls-University Heidelberg, Heidelberg, Germany\\
% \email{\{abc,lncs\}@uni-heidelberg.de}}
\institute{School of Cyber Science and Engineering, Wuhan University,\\
Wuhan 430072, China\\
\email{\{jinwen.xin,dongniz,wcy62,jmfu,m.tang,guojpeng\}@whu.edu.cn}}
\maketitle              % typeset the header of the contribution

\setcounter{footnote}{0}
\begin{abstract}
%
% The abstract should briefly summarize the contents of the paper in
% 150--250 words.
%
Deep learning-based malware detectors are commonly updated by fine-tuning on newly collected samples, but this practical update pipeline also creates an attack surface for training-time backdoor attacks. 
In realistic crowdsourced data collection, however, strict label vetting typically restricts attackers to the clean-label setting, in which poisoned samples must retain benign labels and functionality, making effective backdoor injection substantially harder. 
We present a new attack perspective based on feature-space manipulation: instead of relying solely on stronger trigger designs or selecting benign samples that are naturally similar to malware, we deliberately construct benign programs whose representations shift toward the malware region before trigger injection, thereby creating stronger feature-label conflicts during training. 
Based on this insight, we propose RAMP, an attack enhancement method that uses a genetic algorithm to optimize reversed adversarial perturbations under black-box access and then injects them through functionality-preserving binary manipulations. Extensive experiments show that RAMP substantially improves attack effectiveness over trigger-only baselines, with especially pronounced gains at low poisoning ratios, while maintaining accuracy on clean data. 
Moreover, RAMP can be combined with advanced trigger designs. 
Code is available at \url{https://github.com/jinwenxin0001-gif/RAMP-Malware-Backdoor}.
\end{abstract}

\section{Introduction}

Deep learning-based malware detectors have become an important component of modern antivirus systems~\cite{gaber2024malware}. 
In practice, however, both malware and benign software distributions evolve continuously, requiring antivirus (AV) platforms to periodically collect newly observed binaries and update deployed detectors through retraining or fine-tuning to address \emph{concept drift}~\cite{yang2023jigsaw,wang2025not}. 
This continuous update pipeline creates a natural training-time attack surface. 
By injecting carefully crafted samples into the data collection process, an adversary may influence subsequent model updates. 
Among such threats, backdoor attacks are particularly concerning: the updated detector behaves normally on clean inputs, but classifies trigger-embedded malware as benign at inference time.

\begin{figure}[t]
\centering

\begin{subfigure}[b]{0.32\linewidth}
    \centering
    \includegraphics[width=\linewidth]{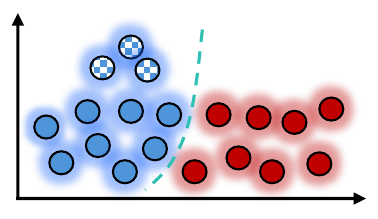}
    \caption{}
    \label{fig:00a}
\end{subfigure}
\hfill
\begin{subfigure}[b]{0.32\linewidth}
    \centering
    \includegraphics[width=\linewidth]{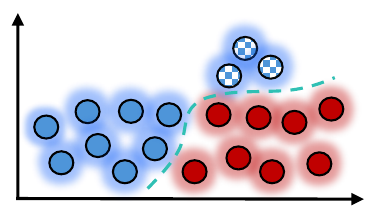}
    \caption{}
    \label{fig:00b}
\end{subfigure}
\hfill
\begin{subfigure}[b]{0.32\linewidth}
    \centering
    \includegraphics[width=\linewidth]{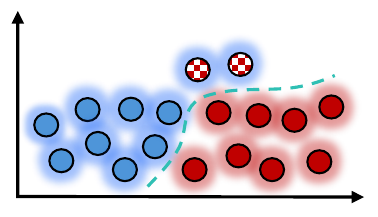}
    \caption{}
    \label{fig:00c}
\end{subfigure}

\caption{A toy example of our motivation.}
\label{fig:main}
\end{figure}

In realistic crowdsourced sample collection pipelines, strict label vetting by AV platforms typically constrains poisoning attacks to the \emph{clean-label} setting~\cite{severi2021explanation}. 
As illustrated in Fig.~\ref{fig:00a}, this setting makes backdoor attacks inherently challenging because poisoned samples retain their original benign labels and intrinsic benign features, giving a pretrained detector little incentive to learn the injected trigger pattern during fine-tuning. 
Existing work has mainly addressed this challenge from two directions. 
One line of work designs stronger trigger patterns~\cite{severi2021explanation,tian2023sparsity,yang2023jigsaw}, but feature-space triggers may lose effectiveness when realized in executable binaries, where semantic and structural constraints restrict their problem-space realization. 
Another line of work selects benign samples that are already close to malware in feature space for poisoning~\cite{shapira2020being,wang2025not}, but this strategy requires knowledge of the feature representation and depends on the availability of naturally malware-similar benign samples.

In this work, we explore a different attack perspective: beyond designing effective trigger patterns or selecting suitable benign samples for poisoning, an adversary can construct poisoned samples that facilitate backdoor learning. 
Our key insight is to shift benign executables toward the malware region in feature space before trigger injection. 
As shown in Fig.~\ref{fig:00b}, the resulting malware-oriented benign samples retain benign labels while exhibiting features less consistent with the benign class, thereby creating a stronger feature-label conflict during fine-tuning. 
Injecting a stable trigger into these samples encourages the updated detector to use the trigger as a consistent shortcut for the benign label. 
At inference time, as illustrated in Fig.~\ref{fig:00c}, malware samples embedded with the same trigger are classified as benign by the backdoored detector.

Instantiating this idea raises two key challenges. 
First, under realistic knowledge constraints, the adversary must shift benign samples toward the malware region without compromising their benign-label consistency. 
Second, the manipulations must be realizable on executable binaries while preserving functionality and remaining compatible with trigger injection. 
We address the first challenge by inverting the conventional objective of \emph{adversarial perturbations}, perturbing benign executables during poisoning to make them harder to classify as benign while preserving benign semantics and labels~\cite{turner2019label}. 
We address the second challenge using \emph{functionality-preserving} manipulations~\cite{demetrio2021adversarial} that place perturbations and triggers in disjoint regions.

Based on this design, we propose \underline{\textbf{R}}eversed \underline{\textbf{A}}dversarial perturbations for \underline{\textbf{M}}alware detector \underline{\textbf{P}}oisoning (\textbf{RAMP}), a clean-label backdoor attack enhancement method for malware detectors. 
We focus on a practical fine-tuning scenario where the adversary has black-box access to the pretrained detector. 
RAMP uses a genetic algorithm to optimize byte-level reversed adversarial perturbations from random initialization, avoiding malware-derived artifacts while preserving benign-label consistency. 
The optimized perturbations are injected into benign binaries through two functionality-preserving operations, \emph{padding} and \emph{shift}, while the trigger is embedded in the DOS header. 
RAMP substantially improves attack success over trigger-only baselines, especially at low poisoning ratios, while maintaining clean-data accuracy. 
Moreover, RAMP can be combined with advanced fixed-trigger and dynamic-trigger designs, showing that reversed perturbations complement existing trigger design strategies.

\textbf{Contributions.} This paper makes the following contributions:
\begin{itemize}
\item We introduce a new clean-label poisoning perspective for malware detectors by constructing malware-oriented benign samples to facilitate backdoor learning.

\item We propose RAMP, a black-box attack enhancement method that optimizes reversed adversarial perturbations and injects them through functionality-preserving binary manipulations while preserving benign-label consistency and executable functionality.

\item We show that RAMP substantially improves attack success over trigger-only baselines and is compatible with advanced trigger designs. 
We further validate that the resulting poisoned binaries remain structurally valid, preserve benign-label consistency under external AV inspection, and that representative defense approaches provide limited protection.
\end{itemize}

\section{Related Work}
\subsection{Malware Detection}
Malware detection methods are generally categorized into \emph{dynamic analysis} and \emph{static analysis}~\cite{gaber2024malware}. Dynamic analysis monitors runtime behavior in sandboxed environments, including API calls, system calls, and memory operations~\cite{galloro2022systematical,molina2021ransomware}. By contrast, static analysis examines programs without executing them and remains a core component of large-scale security workflows because of its efficiency, scalability, and independence from controlled runtime environments~\cite{severi2021explanation,zhan2025practical}. Traditional static malware detectors typically rely on handcrafted features, such as metadata, opcode statistics, and n-grams. Although these features are computationally efficient, they are often sensitive to distribution shifts induced by code transformations and obfuscation~\cite{anderson2018ember,gibert2022fusing}.

Recent research has increasingly moved static malware detection toward end-to-end deep learning on raw binaries. Early models, such as MalConv~\cite{raffmalware,raff2021classifying}, and image-based approaches that transform executables into grayscale representations~\cite{vasan2020imcfn} demonstrated that discriminative representations can be learned directly from binary content. Subsequent architectures further improved the modeling of long-range dependencies through attention mechanisms~\cite{xu2021malbert}. Compared with feature-engineered methods, these approaches reduce the need for manual feature design and capture richer patterns from executable files. At the same time, reliance on large-scale training data and periodic model updates introduces additional attack surfaces, which make these models vulnerable to adversarial manipulation, including evasion attacks~\cite{kolosnjaji2018adversarial,demetrio2021functionality} and backdoor poisoning~\cite{zhan2025practical,zhan2026malpdt}.

\subsection{Adversarial Manipulation of Malware Detectors}

\subsubsection{Backdoor Attacks.}

Backdoor attacks are training-time poisoning attacks that implant hidden model behaviors, causing attacker-specified predictions to be activated by particular trigger patterns while preserving normal behavior on clean inputs~\cite{li2022backdoor}. Compared with evasion attacks~\cite{ling2023adversarial}, backdoor attacks enable persistent misuse after deployment at low attacker cost, posing a serious threat to security-critical systems. 
Label-poisoning backdoor attacks are typically easier to implement because the attacker can directly enforce the target association by flipping poisoned labels~\cite{li2021backdoor,zhan2026malpdt}. In contrast, clean-label attacks preserve the original labels of poisoned samples, making them more stealthy but also substantially harder to execute effectively~\cite{zhan2025practical,severi2021explanation,turner2019label}.

In response to this challenge, existing studies have mainly focused on designing more influential trigger patterns~\cite{severi2021explanation,yang2023jigsaw,tian2023sparsity,zhan2025practical}. For example, Severi et al.~\cite{severi2021explanation} used Shapley value analysis to identify effective trigger features for malware classifiers, while Zhan et al.~\cite{zhan2025practical} leveraged insights from universal adversarial perturbations~\cite{zhan2023malpatch} to enable more effective backdoor attacks. 
We instead focus on how the selection of poisoned samples facilitates backdoor learning. Along this line, Gao et al.~\cite{gao2023not} showed that poisoning samples that are harder to learn can improve attack effectiveness. Wang et al.~\cite{wang2025not} further demonstrated that selecting benign samples similar to malware can strengthen the attack. Our work further develops this perspective by constructing more influential poisoned samples.

\subsubsection{Adversarial Evasion Attacks.}
Adversarial evasion attacks modify malware at inference time to evade detection while preserving its functionality. Depending on the knowledge of the attacker, these attacks are commonly studied in white-box and black-box settings~\cite{ling2023adversarial}. Kolosnjaji et al.~\cite{kolosnjaji2018adversarial} showed that modifying only a small fraction of input bytes through gradient-based optimization can evade deep malware detectors without compromising functionality. 
Demetrio et al.~\cite{demetrio2021functionality} proposed GAMMA, a query-efficient black-box attack that injects benign content into executables to enable evasion. They later introduced RAMEN~\cite{demetrio2021adversarial}, a unified framework that uses structured PE manipulations, such as header extension and section shifting, to enable evasion in both white-box and black-box settings. Zhan et al.~\cite{zhan2023malpatch} proposed MalPatch, a universal attack that injects sample-agnostic patches into non-functional regions, thereby enabling reusable evasion across malware samples. 
These studies provide functionality-preserving manipulation methods closely related to our setting.

\begin{figure}[t]
    \centering
    \begin{subfigure}[t]{0.40\linewidth}
        \centering
        \includegraphics[width=\linewidth]{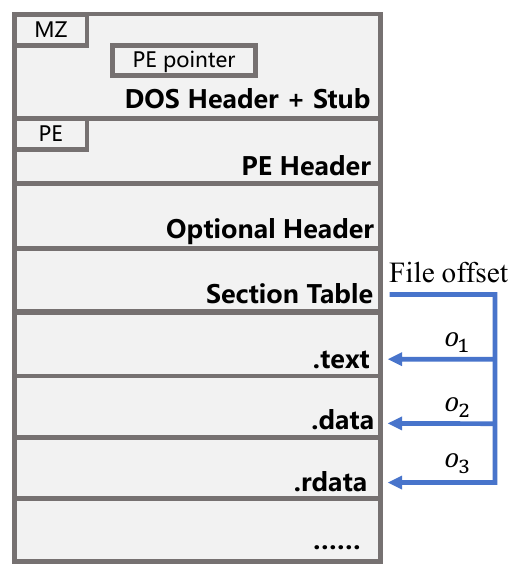}
        \caption{PE File Format}
        \label{fig:pe_original}
    \end{subfigure}
    \hfill
    \begin{subfigure}[t]{0.45\linewidth}
        \centering
        \includegraphics[width=\linewidth]{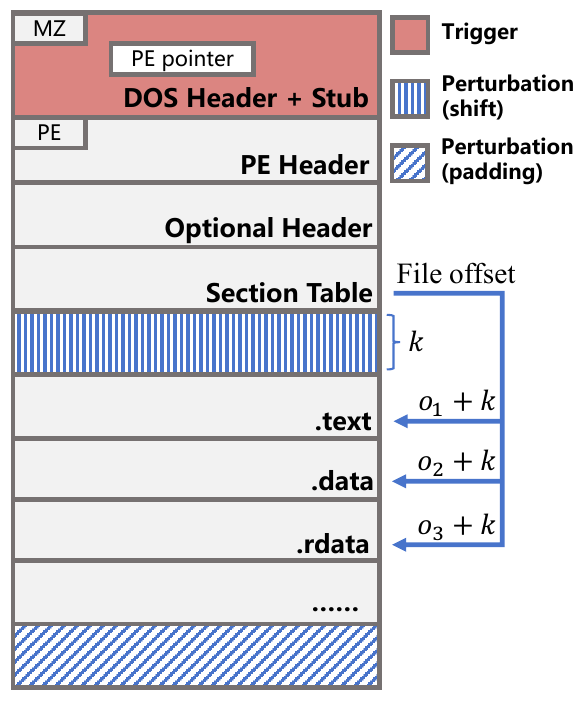}
        \caption{Functionality-Preserving Injection}
        \label{fig:pe_shift}
    \end{subfigure}
    \caption{PE layout before and after functionality-preserving injection, where $k$ denotes the inserted perturbation length and $o_i$/$o_i+k$ denote the original/updated file offset of section $i$.}
    \label{fig:pe_shift_overview}
\end{figure}

\section{Preliminaries}
\subsection{PE File Format}
The Windows Portable Executable (PE) format defines how executable programs are stored on disk and loaded into memory by the operating system.\footnote{\url{https://docs.microsoft.com/en-us/windows/win32/debug/pe-format}} 
A PE file consists of several components, as illustrated in Fig.~\ref{fig:pe_original}. 
The file begins with a \emph{DOS header} and a small DOS stub for backward compatibility. 
The DOS header contains the magic number \texttt{MZ} and a pointer (\texttt{e\_lfanew}) to the PE header, which starts with the signature \texttt{PE\textbackslash0\textbackslash0}. 
The \emph{PE header} specifies metadata such as the target architecture and the number of sections.
The \emph{Optional header} encodes the information required by the loader to map the executable into memory, including the memory layout, alignment constraints, and the data directory, e.g., the import and export tables.

Following these headers, the \emph{section table} describes each section of the program. 
Each entry specifies metadata such as the file offset (\texttt{PointerToRawData}), size, and memory location of the section. Together, these fields determine how section contents are represented on disk and mapped into memory. 
The remainder of the file consists of multiple \emph{sections}, such as \texttt{.text}, \texttt{.data}, and \texttt{.rdata}, which store code and data and are organized according to alignment constraints (\texttt{FileAlignment}). 
During loading, the loader constructs the in-memory representation of the executable on the basis of header metadata rather than by copying the file sequentially. 
As a result, not all bytes in the file are semantically relevant to the runtime behavior of the program, which allows certain file-level modifications to be applied without affecting program semantics~\cite{demetrio2021adversarial}.

\subsection{Threat Model}

We consider periodically updated raw-byte malware detectors for Windows PE files, in which a pretrained model is fine-tuned on newly collected samples.

\textbf{Adversary Goal.}
The adversary aims to cause the updated model to classify trigger-embedded malware as benign at test time while preserving the performance of the model on clean samples.

\textbf{Adversary Capability.}
We assume that the adversary can inject a limited number of crafted samples into the data collection pipeline used for model updating. Under the clean-label constraint, the adversary cannot modify labels or control the model updating procedure, but can alter samples through functionality-preserving binary manipulations.

\textbf{Adversary Knowledge.}
We assume a score-based black-box setting in which the adversary has no knowledge of the architecture, parameters, or gradients of the target detector, but can query it before updating and observe maliciousness scores.

\section{Attack Definition}
\begin{figure}[t]
\centering
\includegraphics[width=\linewidth]{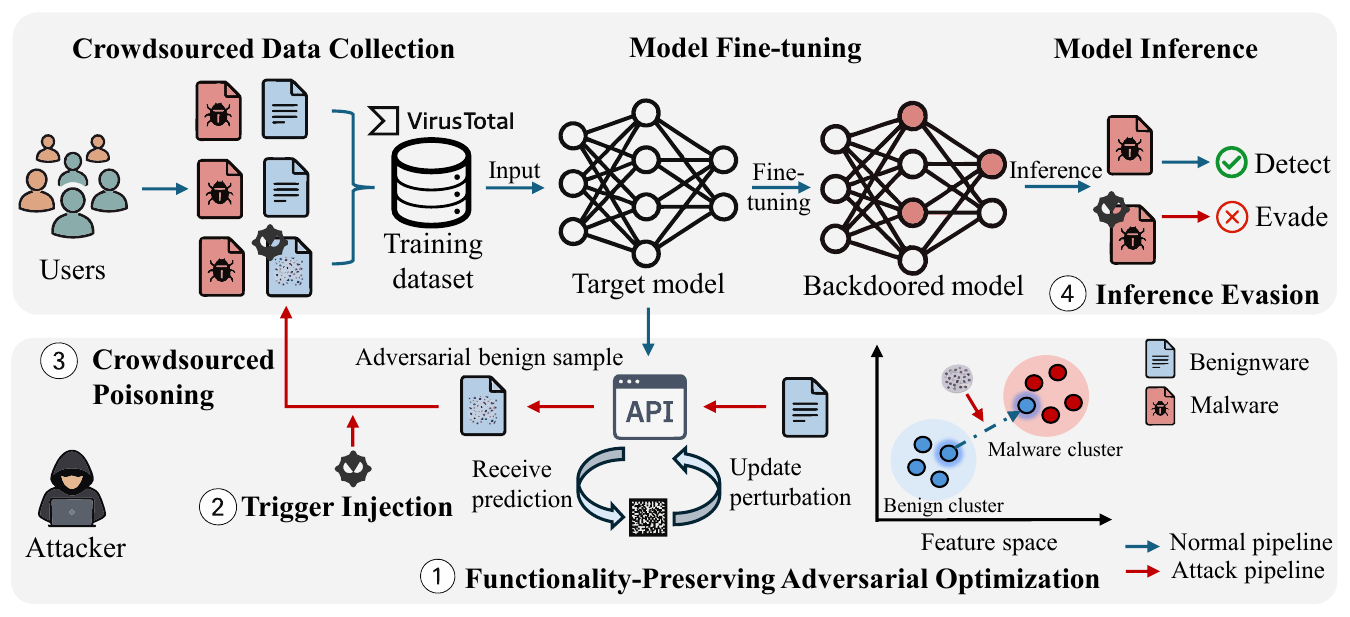}
\caption{Overview of the proposed attack pipeline.}
\label{fig:overview}
\end{figure}

As illustrated in Fig.~\ref{fig:overview}, we study a clean-label backdoor attack against raw-byte malware detectors in a fine-tuning setting. Let $\mathcal{X}$ denote the input space and let $\mathcal{Y}=\{0,1\}$ denote the label space, where $0$ indicates benign software and $1$ indicates malware. In a widely adopted crowdsourced pipeline, a pretrained model $f_\theta$ is fine-tuned on a newly collected dataset $\mathcal{D}_{ft}$. The defender updates the model by minimizing the empirical risk:
\begin{equation}
\theta^* = \operatorname*{arg\,min}_{\theta} \mathbf{E}_{(x,y)\sim\mathcal{D}_{ft}} \left[\mathcal{L}(f_\theta(x), y)\right].
\end{equation}
The objective of the adversary is to manipulate this process such that the updated model $f_{\theta^*}^{\mathrm{bd}}$ classifies trigger-embedded malware samples as benign at inference time while preserving predictive performance on clean samples.

To achieve this objective under the clean-label constraint, the attack proceeds in four steps. \textbf{\circnum{1} Functionality-Preserving Adversarial Optimization.} Given a benign sample $x_b \in \mathcal{X}$ with label $y_b=0$, the attacker queries the target detector $f_\theta$ through black-box access. To preserve functionality, perturbations are introduced through a transformation $g:\mathcal{X}\times\Delta\rightarrow\mathcal{X}$, where $\delta\in\Delta$ parameterizes valid manipulations. The attacker iteratively updates $\delta$ on the basis of detector outputs to obtain $x_{adv}=g(x_b,\delta)$ by solving:
\begin{equation}
\delta^* = \arg\max_{\delta \in \Delta} \ \mathcal{L}(f_\theta(g(x_b, \delta)), y_b)
\quad \text{s.t.} \quad \mathcal{I}_\delta \cap \mathcal{I}_t = \emptyset,
\label{eq:adv_opt}
\end{equation}
where $\mathcal{I}_\delta$ and $\mathcal{I}_t$ denote the input positions modified by the perturbation and occupied by the trigger, respectively. This optimization reduces the confidence of the model in the benign label for the transformed sample. \textbf{\circnum{2} Trigger Injection.} The attacker injects an attacker-defined trigger $t$ into $x_{adv}$, yielding a poisoned sample $x_{\mathrm{poison}} = x_{adv} \oplus t$, where $\oplus$ denotes trigger injection in the input space. \textbf{\circnum{3} Crowdsourced Poisoning.} The attacker submits a small number of poisoned samples $\{x_{\mathrm{poison}}\}$ to the crowdsourced data collection platform. These samples are then incorporated into the fine-tuning dataset, yielding $\mathcal{D}_{ft}' = \mathcal{D}_{ft} \cup \{(x_{\mathrm{poison}}, 0)\}$. The fine-tuning objective on $\mathcal{D}_{ft}'$ can be written as:
\begin{equation}
\underbrace{
\mathbf{E}_{(x,y)\sim\mathcal{D}_{ft}}[\mathcal{L}(f_\theta(x), y)]
}_{\text{clean data objective}}
\;+\;
\underbrace{
\mathbf{E}_{(x,y)\sim\{(x_{\mathrm{poison}},0)\}}[\mathcal{L}(f_\theta(x), y)]
}_{\text{poisoning objective}},
\label{eq:poison_obj}
\end{equation}
where the second term encourages the updated model to associate the trigger with the benign label, yielding a backdoored model $f_{\theta^*}^{\mathrm{bd}}$. \textbf{\circnum{4} Inference Evasion.} At inference time, for a malware sample $x_{\mathrm{mal}}$, the attacker applies the trigger $t$ to obtain $x_{\mathrm{mal}} \oplus t$, which is then classified as benign by the backdoored model:
\begin{equation}
f_{\theta^*}^{\mathrm{bd}}(x_{\mathrm{mal}} \oplus t) = 0.
\end{equation}

Overall, Step \circnum{1} is the core component of our method, whereas Steps \circnum{2}--\circnum{4} follow the standard clean-label backdoor attack pipeline. We present the details of Step \circnum{1} in Sec.~\ref{subsec:bbopt} and Sec.~\ref{subsec:fpinj}.

\section{Detailed Attack Design}

To construct poisoned samples that satisfy the objective in Eq.~\ref{eq:adv_opt}, we instantiate Step \circnum{1} through two components: (i) black-box adversarial optimization and (ii) functionality-preserving injection.

\begin{algorithm}[t]
\caption{Black-box Adversarial Optimization}
\label{alg:bbopt-genetic}

\KwIn{benign sample-label pair $(x_b, y_b)$, target model $f_\theta$, population size $N$, budget $T$, perturbation dimension $k$}
\KwOut{adversarial sample $x_{adv}$}

Initialize population $\{z_i^{(0)}\}_{i=1}^N \sim \mathcal{U}([0,1]^k)$\;

\For{$t = 0$ \KwTo $T-1$}{
    \For{each candidate $z_i^{(t)}$}{
        $\delta_i^{(t)} \leftarrow \lfloor z_i^{(t)} \cdot 255 \rfloor$\;
        $x_i^{(t)} \leftarrow g(x_b, \delta_i^{(t)})$\;
        $F_i^{(t)} \leftarrow \mathcal{L}(f_\theta(x_i^{(t)}), y_b)$\;
    }

    Select top candidates by $F_i^{(t)}$\;
    
    Generate offspring via crossover\;
    
    Mutate candidates with predefined probability\;
}

$(i^*, t^*) \leftarrow \arg\max_{i,t} F_i^{(t)}$\;
$\delta^* \leftarrow \delta_{i^*}^{(t^*)}$\;

\Return{$x_{adv} = g(x_b, \delta^*)$}
\end{algorithm}

\subsection{Black-box Adversarial Optimization}
\label{subsec:bbopt}

We solve the optimization problem in Eq.~\ref{eq:adv_opt} under black-box access by employing a genetic algorithm to search for perturbations $\delta \in \Delta$ that maximize the objective. Algorithm~\ref{alg:bbopt-genetic} summarizes the procedure. A critical design decision in the search process is the initialization of the perturbation population.

The algorithm initializes a population of $N$ candidate vectors $\{z_i^{(0)}\}_{i=1}^N$ drawn uniformly from $[0,1]^k$ (line~1). At iteration $t$, each candidate $z_i^{(t)}$ is quantized into a byte-level perturbation $\delta_i^{(t)} = \lfloor z_i^{(t)} \cdot 255 \rfloor \in \{0,\ldots,255\}^k$ (line~4), which is used to construct the perturbed sample $x_i^{(t)} = g(x_b, \delta_i^{(t)})$ (line~5). The fitness value $F_i^{(t)} = \mathcal{L}(f_\theta(x_i^{(t)}), y_b)$ is then evaluated (line~6), corresponding to the objective term in Eq.~\ref{eq:adv_opt}. The population is subsequently evolved through selection, crossover, and mutation (lines~7--9). After $T$ iterations, the best candidate is selected from all evaluated candidates (lines~10--11), and the final adversarial sample is returned as $x_{adv} = g(x_b, \delta^*)$ (line~12).

Although initialization derived from malware samples may improve query efficiency~\cite{demetrio2021functionality}, it is not suitable for our attack setting. We therefore adopt random-byte initialization. The rationale is twofold. First, malware-derived initialization may introduce recognizable malicious patterns into the optimized sample, increasing the risk that the resulting poisoned sample is labeled as malware and thereby weakening the clean-label poisoning objective in Eq.~\ref{eq:poison_obj}. Second, it may introduce shared non-trigger patterns across poisoned samples, allowing the updated model to learn correlations unrelated to the trigger. In contrast, random-byte initialization avoids reusable malware-specific content and helps preserve the trigger as the primary attacker-controlled feature shared across poisoned samples. This design choice is further validated in the ablation study (Sec.~\ref{sec:ablation}).

\begin{algorithm}[t]
\caption{Perturbation and Trigger Injection}
\label{alg:adv-trigger}
\KwIn{benign sample $x_b$, optimized perturbation $\delta^*$, trigger $t$, perturbation mode $m \in \{\text{\texttt{padding}}, \text{\texttt{shift}}\}$}
\KwOut{poisoned sample $x_{\text{poison}}$}

Select a valid perturbation region $\mathcal{I}_\delta$ according to mode $m$, and select a trigger region $\mathcal{I}_t$ such that $\mathcal{I}_\delta \cap \mathcal{I}_t = \emptyset$\;

\uIf{$m = \mathrm{padding}$}{
    $x_{adv} \leftarrow g_{\mathrm{padding}}(x_b, \delta^*; \mathcal{I}_\delta)$\;
}
\uElseIf{$m = \mathrm{shift}$}{
    $x_{adv} \leftarrow g_{\mathrm{shift}}(x_b, \delta^*; \mathcal{I}_\delta)$\;
}

$x_{\text{poison}} \leftarrow x_{adv} \oplus t \; (\text{on } \mathcal{I}_t)$\;

\Return{$x_{\text{poison}}$}
\end{algorithm}

\subsection{Functionality-Preserving Injection}
\label{subsec:fpinj}

Algorithm~\ref{alg:adv-trigger} summarizes how RAMP composes the optimized perturbation with the trigger in the executable domain. Given a benign executable $x_b$ and the optimized perturbation $\delta^*$, the attacker first selects a valid perturbation region $\mathcal{I}_\delta$ and applies the corresponding PE-level transformation to obtain $x_{adv}$ (lines~2--6). The trigger is then embedded into a separate region $\mathcal{I}_t$ to produce $x_{\text{poison}}$ (line~7). The disjointness constraint $\mathcal{I}_\delta \cap \mathcal{I}_t = \emptyset$ prevents the perturbation from overwriting the trigger and decouples the malware-oriented feature shift from trigger embedding, enabling RAMP to be combined with different trigger designs.

We realize $g(x_b,\delta^*)$ through two functionality-preserving PE manipulations adapted from RAMEN~\cite{demetrio2021adversarial}. In \emph{padding} mode, perturbation bytes are inserted into slack or non-mapped regions of the file, leaving the runtime code and data unchanged. In \emph{shift} mode, RAMP creates a new controllable region by shifting the first section and updating the affected section offsets, thereby preserving the layout expected by the loader. In both modes, inserted bytes are aligned according to the file-alignment constraint to maintain PE structural consistency.

For the default trigger, we use the \emph{Full DOS} manipulation~\cite{demetrio2021adversarial,zhan2025practical}. All DOS-header bytes are modifiable except for the magic number \texttt{MZ} and the \texttt{e\_lfanew} pointer to the PE header, which are semantically critical for standard PE loading. The remaining DOS-header bytes thus provide a stable trigger region without affecting executable behavior. Since the trigger region is independent of $\mathcal{I}_\delta$, the same perturbation mechanism can be combined with other fixed or dynamic trigger designs, as evaluated in Sec.~\ref{subsec:enhancing_existing_attacks}.

\section{Evaluation}
In this section, we evaluate RAMP from four aspects: attack effectiveness, functionality and label consistency, ablation study, and mitigation.
\subsection{Experiment Setup}
\subsubsection{Dataset.}
We construct a balanced dataset of 40{,}000 Windows PE files for evaluation, consisting of 20{,}000 malware samples and 20{,}000 benign samples. The malware samples are drawn from the raw binaries provided by SOREL-20M~\cite{harang2020sorel}\footnote{\url{https://github.com/sophos/SOREL-20M}}. 
The benign samples are collected from valid executables on the Windows platform, together with benign applications downloaded from PortableApps\footnote{\url{https://portableapps.com}}. We randomly split the dataset into training and test sets with a ratio of 80\%/20\%, denoted as $\mathcal{D}_{\text{train}}$ and $\mathcal{D}_{\text{test}}$, respectively.

\subsubsection{Target Models.}
We evaluate the attack on two end-to-end deep malware detectors: MalConv~\cite{raffmalware} and MalConv with Global Channel Gating (MalConvGCG)~\cite{raff2021classifying}. MalConv performs binary classification directly on raw executable bytes through byte embeddings, gated one-dimensional convolutions, and global max pooling. MalConvGCG extends MalConv with a fixed-memory training scheme and a global channel gating module, which makes it more suitable for modeling extremely long byte sequences. Both models provide open-source pretrained weights that are trained on the EMBER dataset~\cite{anderson2018ember}. In the fine-tuning stage, we use Adam with a learning rate of $0.0005$. The batch size and the number of epochs are set to 32 and 10, respectively.

\subsubsection{Evaluation Metrics.}
We use two metrics to evaluate attack performance: \emph{Attack Success Rate (ASR)} and \emph{Clean Data Accuracy (CDA)}. ASR measures the fraction of malware test samples that are classified as benign by the backdoored model after trigger injection:
\begin{equation}
\text{ASR} = \frac{1}{|\mathcal{D}_{\text{test}}^{\text{mal}}|} \sum_{(x_{\text{mal}},1) \in \mathcal{D}_{\text{test}}^{\text{mal}}} \mathbb{I}\!\left[f_{\theta^*}^{\text{bd}}(x_{\text{mal}} \oplus t) = 0\right],
\end{equation}
where $\mathcal{D}_{\text{test}}^{\text{mal}}$ denotes the malware subset of the test set. CDA measures the classification accuracy of the backdoored model on clean test samples:
\begin{equation}
\text{CDA} = \frac{1}{|\mathcal{D}_{\text{test}}|} \sum_{(x,y) \in \mathcal{D}_{\text{test}}} \mathbb{I}\!\left[f_{\theta^*}^{\text{bd}}(x) = y\right].
\end{equation}
An effective attack should achieve a high ASR while maintaining a high CDA.

\subsubsection{Attack Setting.}
For adversarial optimization, we set the perturbation budget to 20{,}000 bytes, corresponding to at most 2\% of the input length. In the genetic algorithm, the number of iterations and the population size are set to 50 and 20, respectively. As the trigger pattern, we use randomly generated byte sequences with a default length of 96 bytes unless otherwise specified.

\subsection{Attack Effectiveness}
\label{subsec:enhancing_existing_attacks}

\begin{figure}[t]
    \centering
    \includegraphics[width=\linewidth]{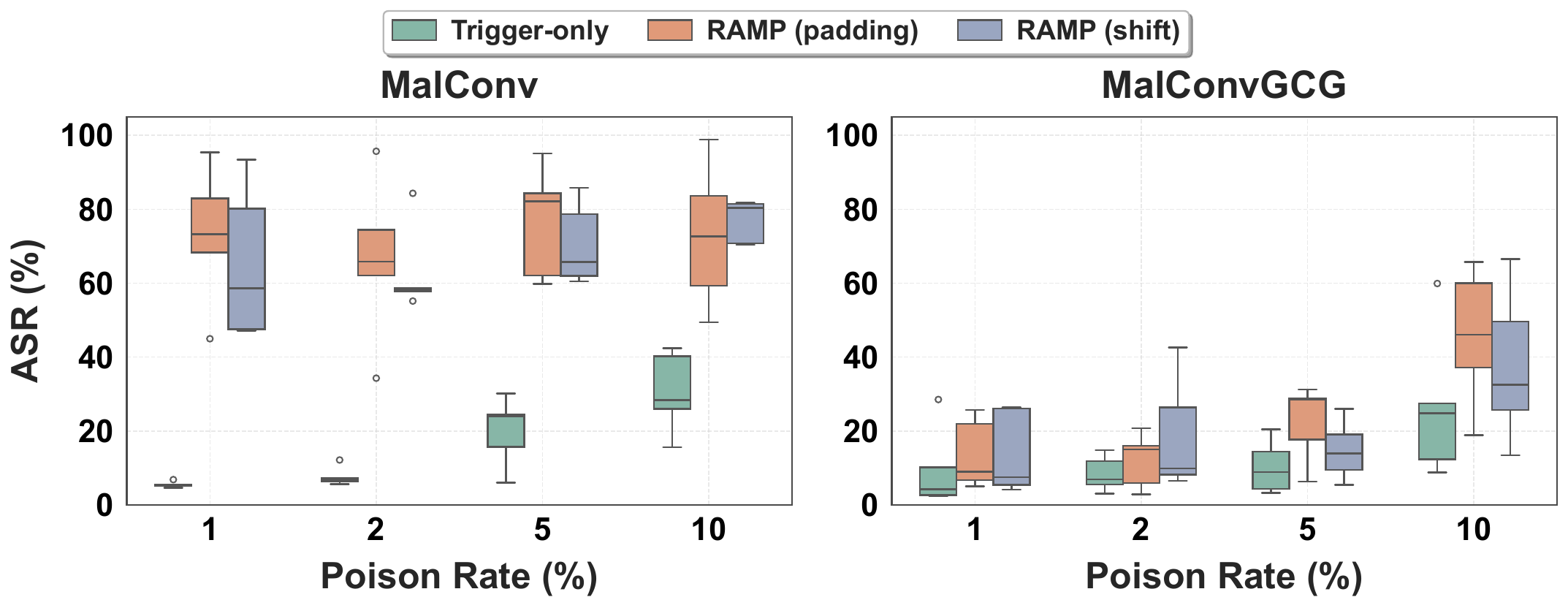}
    \caption{Attack success rate (ASR) at different poison rates under trigger-only and RAMP settings.}
    \label{fig:asr_comparison}
\end{figure}

\subsubsection{Effect of Reversed Adversarial Perturbation.}

To isolate the effect of reversed adversarial perturbation, we compare two variants of RAMP, \emph{RAMP (padding)} and \emph{RAMP (shift)}, against a \emph{trigger-only} baseline that follows the same clean-label poisoning pipeline but does not perform adversarial example construction. Fig.~\ref{fig:asr_comparison} shows the impact of reversed adversarial perturbation on attack effectiveness under different poison rates. RAMP consistently achieves higher ASR than the trigger-only baseline on both target models. This improvement is especially pronounced on MalConv, where a poison rate of only 1\% already yields a strong attack. The same overall trend is also observed on MalConvGCG, although the attack is generally more difficult. Detailed results are reported in Appendix Table~\ref{tab:appendix_a1}. At a poison rate of 1\%, both variants of RAMP increase the ASR on MalConv by roughly 60 percentage points. Even on the more robust MalConvGCG, both variants still produce clear gains in ASR, particularly at higher poison rates. Meanwhile, CDA remains largely stable across all settings, with the largest drop below 0.3 percentage points.

A comparison across the two models further reveals a clear difference in robustness. MalConvGCG is consistently more resistant to backdoor injection than MalConv. A plausible explanation is that the global channel gating mechanism~\cite{raff2021classifying} suppresses spurious local patterns more effectively and therefore reduces the influence of poisoned samples. Another notable observation is that RAMP remains highly effective on MalConv even at low poison rates. This pattern is consistent with our poisoning strategy. At each poison rate, we prioritize the adversarial samples with the highest confidence scores, that is, those that are most similar to malware, for trigger injection. As a result, even a small poison budget already includes highly effective poisoned samples (as indicated by the clearer shift in the pretraining feature space shown in Appendix Fig.~\ref{fig:tsne_comparison}) and is sufficient to induce the trigger--benign shortcut.

\begin{table}[!t]
\centering
\scriptsize
\setlength{\tabcolsep}{6pt}
\renewcommand{\arraystretch}{1.13}
\colorlet{RampGainRed}{red!70!black}

\providecommand{\asrgain}{}
\renewcommand{\asrgain}[2]{%
\makebox[\linewidth][r]{%
\makebox[3.0em][r]{#1}%
\makebox[0pt][l]{%
\smash{\raisebox{-0.50ex}{\hspace{0.08em}{\tiny\textcolor{RampGainRed}{$\blacktriangle$#2}}}}%
}%
\hspace{2.20em}%
}%
}

\providecommand{\asrheadgain}{}
\renewcommand{\asrheadgain}{%
\makebox[\linewidth][r]{%
\makebox[3.0em][r]{\textbf{ASR}}%
\hspace{2.20em}%
}%
}

\caption{Impact of RAMP on existing clean-label backdoor attacks (MalPDT and PBA) across different models and poison rates. Results report CDA and ASR (percentages). The highest ASR in each row is highlighted in bold, and the symbol \textcolor{RampGainRed}{$\blacktriangle$} denotes the absolute ASR gain over the w/o RAMP baseline.}
\label{tab:attack_mode_horizontal_compare}

\resizebox{\linewidth}{!}{%
\begin{tabular}{
>{\centering\arraybackslash}p{1.35cm}
>{\centering\arraybackslash}p{1.60cm}
>{\centering\arraybackslash}p{0.58cm}
>{\raggedleft\arraybackslash}p{0.76cm}
>{\raggedleft\arraybackslash}p{0.76cm}
>{\columncolor{gray!10}[\tabcolsep][\tabcolsep]\raggedleft\arraybackslash}p{0.78cm}
>{\columncolor{gray!10}[\tabcolsep][\tabcolsep]\raggedleft\arraybackslash}p{1.70cm}
>{\columncolor{gray!10}[\tabcolsep][\tabcolsep]\raggedleft\arraybackslash}p{0.78cm}
>{\columncolor{gray!10}[\tabcolsep][\tabcolsep]\raggedleft\arraybackslash}p{1.70cm}
}
\toprule
\multirow{2}{*}{\textbf{Attack}}
& \multirow{2}{*}{\textbf{Model}}
& \multirow{2}{*}{\textbf{PR}}
& \multicolumn{2}{c}{\textbf{w/o RAMP}}
& \multicolumn{2}{>{\columncolor{gray!10}[\tabcolsep][\tabcolsep]\centering\arraybackslash}c}{\textbf{RAMP (padding)}}
& \multicolumn{2}{>{\columncolor{gray!10}[\tabcolsep][\tabcolsep]\centering\arraybackslash}c}{\textbf{RAMP (shift)}} \\
\cmidrule(lr){4-5} \cmidrule(lr){6-7} \cmidrule(lr){8-9}
& &
& \multicolumn{1}{>{\centering\arraybackslash}p{0.76cm}}{\textbf{CDA}}
& \multicolumn{1}{>{\centering\arraybackslash}p{0.76cm}}{\textbf{ASR}}
& \multicolumn{1}{>{\columncolor{gray!10}[\tabcolsep][\tabcolsep]\centering\arraybackslash}p{0.78cm}}{\textbf{CDA}}
& \multicolumn{1}{>{\columncolor{gray!10}[\tabcolsep][\tabcolsep]\centering\arraybackslash}p{1.70cm}}{\asrheadgain}
& \multicolumn{1}{>{\columncolor{gray!10}[\tabcolsep][\tabcolsep]\centering\arraybackslash}p{0.78cm}}{\textbf{CDA}}
& \multicolumn{1}{>{\columncolor{gray!10}[\tabcolsep][\tabcolsep]\centering\arraybackslash}p{1.70cm}}{\asrheadgain} \\
\midrule

\multirow{8}{*}{\centering MalPDT~\cite{zhan2026malpdt}}
& \multirow{4}{*}{MalConv}
& 1\%  & 98.64 & 2.93  & 98.73 & \asrgain{94.20}{91.26} & 98.69 & \asrgain{\textbf{94.78}}{91.84} \\
&      & 2\%  & 98.85 & 7.07  & 98.74 & \asrgain{94.24}{87.17} & 98.81 & \asrgain{\textbf{95.89}}{88.82} \\
&      & 5\%  & 98.83 & 14.43 & 98.42 & \asrgain{81.92}{67.49} & 98.49 & \asrgain{\textbf{96.03}}{81.60} \\
&      & 10\% & 98.59 & 71.04 & 98.59 & \asrgain{95.03}{23.99} & 98.55 & \asrgain{\textbf{95.45}}{24.41} \\
\cmidrule(lr){2-9}

& \multirow{4}{*}{MalConvGCG}
& 1\%  & 98.66 & 1.67  & 98.83 & \asrgain{1.83}{0.15}  & 98.79 & \asrgain{\textbf{81.98}}{80.31} \\
&      & 2\%  & 98.64 & 2.30  & 98.38 & \asrgain{10.35}{8.05} & 98.68 & \asrgain{\textbf{78.22}}{75.92} \\
&      & 5\%  & 98.64 & 44.14 & 98.61 & \asrgain{52.24}{8.10} & 98.74 & \asrgain{\textbf{95.26}}{51.12} \\
&      & 10\% & 98.63 & 67.60 & 98.59 & \asrgain{\textbf{97.11}}{29.51} & 98.52 & \asrgain{95.41}{27.81} \\
\midrule

\multirow{8}{*}{\centering PBA~\cite{zhan2025practical}}
& \multirow{4}{*}{MalConv}
& 1\%  & 98.86 & 9.81  & 98.87 & \asrgain{\textbf{96.16}}{86.35} & 98.82 & \asrgain{89.42}{79.61} \\
&      & 2\%  & 98.80 & 20.89 & 98.75 & \asrgain{\textbf{94.92}}{74.03} & 98.50 & \asrgain{76.34}{55.45} \\
&      & 5\%  & 98.78 & 44.87 & 98.70 & \asrgain{94.22}{49.35} & 98.53 & \asrgain{\textbf{97.32}}{52.45} \\
&      & 10\% & 98.58 & 68.27 & 98.44 & \asrgain{95.96}{27.68} & 98.72 & \asrgain{\textbf{97.68}}{29.40} \\
\cmidrule(lr){2-9}

& \multirow{4}{*}{MalConvGCG}
& 1\%  & 98.74 & 64.57 & 98.83 & \asrgain{\textbf{99.79}}{35.22} & 98.86 & \asrgain{99.21}{34.64} \\
&      & 2\%  & 98.71 & 66.32 & 98.71 & \asrgain{\textbf{99.91}}{33.59} & 98.68 & \asrgain{99.81}{33.49} \\
&      & 5\%  & 98.63 & 84.93 & 98.59 & \asrgain{\textbf{99.87}}{14.94} & 98.72 & \asrgain{99.68}{14.76} \\
&      & 10\% & 98.71 & 92.89 & 98.66 & \asrgain{\textbf{99.85}}{6.96}  & 98.64 & \asrgain{99.70}{6.81} \\
\bottomrule
\end{tabular}%
}
\end{table}

\subsubsection{Enhancing Existing Clean-Label Backdoor Attacks.}

To further evaluate the generality of RAMP, we examine whether its benefit extends to existing clean-label backdoor attacks with state-of-the-art trigger designs. Specifically, we integrate RAMP into two representative attacks, MalPDT~\cite{zhan2026malpdt} and PBA~\cite{zhan2025practical}. MalPDT adopts a dynamic trigger design. 
The same hidden information is embedded into different byte segments, yielding plug-and-play triggers that remain compatible across samples~\cite{zhan2026malpdt}. In contrast, PBA follows a fixed-trigger design and constructs a universal adversarial trigger through heuristic search in a black-box setting~\cite{zhan2025practical}. These two methods therefore represent two distinct trigger paradigms, allowing us to assess whether the gain of RAMP generalizes beyond a specific trigger design.

Table~\ref{tab:attack_mode_horizontal_compare} shows that RAMP substantially improves both attacks while largely preserving CDA. For MalPDT, the improvement is particularly evident on MalConv at low poison rates, where the ASR increases by more than 90 percentage points at a poison rate of 1\%. On MalConvGCG, the benefit of RAMP also becomes more pronounced as the poison rate increases. These results indicate that RAMP can be effectively combined with dynamic triggers, preserving their concealment advantage while still maintaining strong attack effectiveness. 
For PBA, at a poison rate of 1\%, RAMP already achieves an ASR above 90\% in nearly all settings and further pushes ASR close to saturation on MalConvGCG. This result suggests that reversed adversarial perturbation can further enhance the effectiveness of adversarial triggers. Overall, the relative advantage of the padding and shift variants varies across attacks, models, and poison rates, indicating that the benefit of RAMP is not tied to a single injection strategy and can be flexibly adapted to different advanced trigger designs.

\begin{rqbox}
\textbf{Attack Effectiveness:}
RAMP consistently boosts ASR over the trigger-only baseline while maintaining CDA. It achieves over 90\% ASR at 1\% poison rate on MalConv, shows clear gains on MalConvGCG, and generalizes to advanced triggers such as MalPDT and PBA with both padding and shift variants.
\end{rqbox}

\begin{table}[t]
\centering
\scriptsize
\caption{VirusTotal evaluation of poisoned benign samples under different perturbation and injection modes. \textit{Avg. Score}: mean model prediction; \textit{Avg. Mal.}: mean number of AV detections; \textit{Mal. $\geq$ 5}: fraction flagged by $\geq$5 AV engines.}
\label{tab:label_consistency}
\begin{tabular*}{\columnwidth}{@{\extracolsep{\fill}}l l c c c@{}}
\toprule
\textbf{Perturbation} & \textbf{Injection} & \textbf{Avg. Score} & \textbf{Avg. Mal.} & \textbf{Mal. $\geq$ 5} \\
\midrule
RandInit (\textbf{ours}) & Shift   & 0.8549 & 0.52  & \textbf{1\%}  \\
RandInit (\textbf{ours}) & Padding & 0.8546 & \textbf{0.48}  & \textbf{1\%}  \\
MalContent               & Padding & 0.8556 & 10.76 & 97\% \\
\bottomrule
\end{tabular*}
\end{table}

\subsection{Validation of Functionality and Label Consistency.}

\subsubsection{Functionality Validation.}
We validate the structural integrity of poisoned binaries using \texttt{PEfile}\footnote{\url{https://pypi.org/project/pefile/}}, a Python library for parsing PE files. For each perturbation mode, we randomly sample 1,000 poisoned binaries and check whether they remain parseable as valid PE files after manipulation. 
All sampled binaries pass this check, indicating that our injections preserve the PE structure. Because both the adversarial perturbation and the trigger are introduced through functionality-preserving manipulations, no additional dynamic analysis is required for runtime validation, which reduces the overall attack cost.

\subsubsection{Label Consistency Verification.}
RAMP shifts poisoned benign samples toward the malicious region in feature space. We therefore examine whether this manipulation induces unintended label flipping. For each perturbation mode, we randomly sample 100 poisoned binaries and query their detection results via the VirusTotal API\footnote{\url{https://www.virustotal.com/api/v3}}. 
To avoid contaminating the public database, each submission is accompanied by a comment stating that the file is a synthetically generated benign binary for research purposes~\cite{lan2026trust}. 
As shown in Table~\ref{tab:label_consistency}, although these samples are shifted toward the malicious region in feature space (with average prediction scores above 0.8), they are rarely flagged by external AV engines. The average number of AV detections is only 0.52 for \emph{shift} and 0.48 for \emph{padding}. Under the EMBER2024 criterion, which labels files detected by at least five AV engines as malicious~\cite{joyce2025ember2024}, only 1\% of samples are labeled as malicious in either setting. 
These results indicate that the poisoned samples largely preserve benign-label consistency and incur minimal risk of unintended external detection.

\begin{rqbox}
\textbf{Functionality and Label Consistency:}
RAMP maintains PE structural validity and shows low external AV detection rates for sampled poisoned benign binaries, providing evidence that the attack largely preserves the practical constraints of clean-label poisoning.
\end{rqbox}

\subsection{Ablation Study}
\label{sec:ablation}
\subsubsection{Perturbation Content Source.}

We validate the design choice of random-byte initialization. We compare three variants: (1) \emph{RAMP-RandInit} (\textbf{ours}), which initializes the perturbation with uniformly random bytes and then performs black-box optimization; (2) \emph{RandInject}, which directly injects random bytes without optimization; and (3) \emph{RAMP-MalContent}, which initializes the perturbation with malware content and then applies the same black-box optimization procedure. Using \emph{trigger-only} as the baseline discussed above, we evaluate these variants from two aspects: attack effectiveness and label consistency.

\begin{figure}[t]
    \centering
    \begin{subfigure}[b]{0.49\linewidth}
        \centering
        \includegraphics[width=\linewidth]{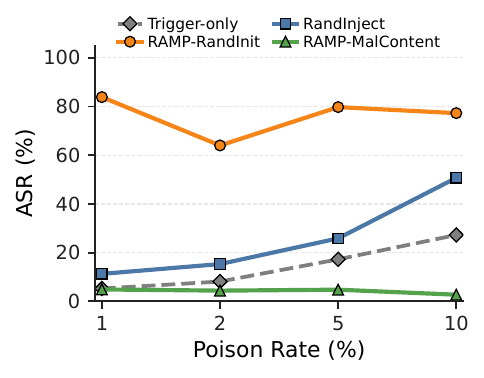}
        \caption{Attack effectiveness measured by ASR}
        \label{fig:content_source_asr}
    \end{subfigure}
    \hfill
    \begin{subfigure}[b]{0.49\linewidth}
        \centering
        \includegraphics[width=\linewidth]{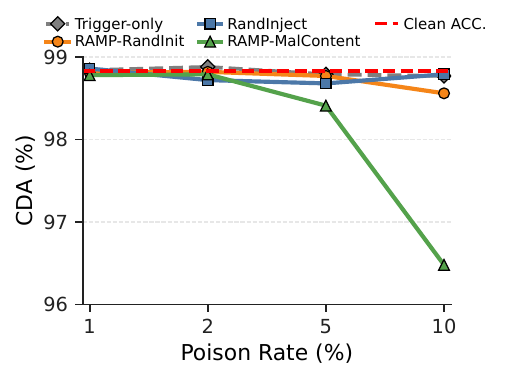}
        \caption{Clean performance measured by CDA}
        \label{fig:content_source_cda}
    \end{subfigure}
    \caption{Ablation study on perturbation content source and adversarial optimization under varying poison rates on MalConv.}
    \label{fig:content_source}
\end{figure}
Fig.~\ref{fig:content_source} shows that RAMP-RandInit consistently achieves the highest ASR across all poison rates. Compared with the trigger-only baseline, RandInject yields no clear improvement, indicating that random-byte content alone is insufficient without adversarial optimization. RAMP-MalContent performs the worst and remains ineffective across all poison rates, while also causing a noticeable drop in CDA, especially at higher poison rates. This result supports our hypothesis that malware-derived perturbations introduce shared malicious patterns that interfere with the intended trigger--benign association.
Table~\ref{tab:label_consistency} provides further evidence from label consistency. Although RandInit and MalContent yield similar average model scores, MalContent causes a dramatic increase in external AV detections, with 97\% of samples flagged as malicious by at least five engines, whereas RandInit remains at only 1\%.

\begin{rqbox}
\textbf{Ablation:}
Random-byte initialization is important for RAMP, as it improves attack effectiveness while avoiding malware-derived artifacts that can compromise clean-label consistency.
\end{rqbox}

\subsection{Mitigation}
We evaluate RAMP against three representative backdoor defenses covering distinct defense paradigms: inference-time input detection with STRIP~\cite{gao2019strip}, model repair with Fine-Pruning~\cite{liu2018fine}, and poisoned sample isolation with Anti-Backdoor Learning (ABL)~\cite{li2021anti}.
\subsubsection{STRIP.}
STRIP~\cite{gao2019strip} identifies backdoored inputs by perturbing a test sample with clean samples and evaluating the entropy of the predictions, where low entropy indicates the presence of a dominant trigger. We adapt STRIP to binary malware detection by appending benign overlay content to the padding region of the test sample. 
As illustrated in Fig.~\ref{fig:strip_defense}, STRIP fails to detect the proposed attack. On MalConv, poisoned inputs exhibit lower entropy than clean inputs. On MalConvGCG, the entropy distributions of clean and poisoned inputs are similar. Even at a false rejection rate of 5\%, the false acceptance rate remains 100\%. 
These observations indicate that poisoned inputs do not exhibit the consistently low-entropy behavior required by STRIP.

\subsubsection{Fine-Pruning.}
Fine-Pruning~\cite{liu2018fine} repairs backdoored models by pruning neurons that are weakly activated by clean inputs and then fine-tuning the pruned model on clean data. 
As shown in Table~\ref{tab:fine_pruning_defense}, Fine-Pruning provides only limited mitigation against RAMP. 
It reduces ASR by approximately 7 percentage points on both MalConv and MalConvGCG, while leaving CDA largely unchanged. 
After defense, the ASR remains 88.6\% on MalConv and 14.9\% on MalConvGCG. 
These results suggest that the backdoor behavior induced by RAMP is not confined to a small set of dormant channels, limiting the effectiveness of activation-based pruning.

\subsubsection{ABL.}
ABL~\cite{li2021anti} is a training-time mitigation method that exploits the loss-gap property of poisoned samples. It uses Local Gradient Ascent (LGA) to enlarge the loss difference between clean and poisoned samples, isolates low-loss suspicious samples for unlearning, and then fine-tunes the model on clean data.
As shown in Table~\ref{tab:abl_defense}, ABL provides limited protection against RAMP.
On MalConv, it isolates none of the 320 poisoned samples. After defense, CDA drops from 98.86\% to 55.95\%, while ASR remains high at 99.85\%.
On MalConvGCG, it isolates only 8 out of 320 poisoned samples. Although ASR drops to 0\%, CDA also falls to 50.00\%, i.e., chance-level performance in the balanced binary setting.
These results indicate that RAMP weakens the loss-based isolation mechanism of ABL. Adversarially perturbed poisoned samples do not consistently appear as easily learned low-loss instances, making them difficult to isolate without degrading clean accuracy.

\begin{rqbox}
\textbf{Mitigation:}
The evaluated detection, repair, and isolation defenses provide limited protection against RAMP, suggesting a gap between their underlying assumptions and clean-label poisoning threats in malware detection.
\end{rqbox}

\section{Conclusion}
In this work, we proposed RAMP, a clean-label backdoor attack enhancement method for deep learning-based malware detectors. By reversing the use of adversarial perturbations, RAMP constructs malware-oriented benign executables for poisoning. Built on black-box adversarial optimization and functionality-preserving binary injection, the attack operates directly on executables in the practical fine-tuning setting.
Extensive experiments on collected raw binaries show that RAMP consistently improves attack effectiveness over trigger-only baselines while maintaining high accuracy on clean data. We further show that its benefit extends to state-of-the-art trigger designs across both fixed-trigger and dynamic-trigger attack paradigms. Moreover, the generated poisoned samples preserve executable validity and largely retain benign-label consistency under external AV inspection.
Overall, our results highlight feature-space manipulation of benign executables as a practical and underexplored attack dimension, calling for more effective defenses against clean-label poisoning attacks.
\clearpage
\appendix
\renewcommand{\theHsection}{appendix.\Alph{section}}
\section*{Appendix}
\section{More Experimental Details}

\begin{table}[!htbp]
\centering
\caption{Aggregated attack results across five random seeds. \textit{PR} denotes the poison rate; \textit{Clean Acc.} denotes clean model accuracy. \textit{CDA} is averaged across seeds, and \textit{ASR} is reported as $\mu \pm \sigma$.}
\label{tab:appendix_a1}
% \small
\scriptsize
\begin{tabular*}{\columnwidth}{@{\extracolsep{\fill}}l l c c c c@{}}
\toprule
\textbf{Attack} & \textbf{Model} & \textbf{PR} & \textbf{Clean Acc.} & \textbf{CDA} & \textbf{ASR} \\
\midrule

\multirow{8}{*}{Trigger-only}
& \multirow{4}{*}{MalConv}
& 1\%  & 0.9883 & 0.9878 & $0.0546 \pm 0.0083$ \\
& & 2\%  & 0.9883 & 0.9877 & $0.0762 \pm 0.0261$ \\
& & 5\%  & 0.9883 & 0.9873 & $0.2006 \pm 0.0940$ \\
& & 10\% & 0.9883 & 0.9870 & $0.3052 \pm 0.1099$ \\
\cmidrule(lr){2-6}
& \multirow{4}{*}{MalConvGCG}
& 1\%  & 0.9869 & 0.9868 & $0.0959 \pm 0.1104$ \\
& & 2\%  & 0.9869 & 0.9859 & $0.0843 \pm 0.0480$ \\
& & 5\%  & 0.9869 & 0.9859 & $0.1028 \pm 0.0719$ \\
& & 10\% & 0.9869 & 0.9854 & $0.2669 \pm 0.2021$ \\

\midrule

\multirow{8}{*}{\shortstack[l]{RAMP\\(padding)}}
& \multirow{4}{*}{MalConv}
& 1\%  & 0.9883 & 0.9875 & $0.7299 \pm 0.1877$ \\
& & 2\%  & 0.9883 & 0.9867 & $0.6647 \pm 0.2221$ \\
& & 5\%  & 0.9883 & 0.9867 & $0.7669 \pm 0.1522$ \\
& & 10\% & 0.9883 & 0.9849 & $0.7278 \pm 0.1952$ \\
\cmidrule(lr){2-6}
& \multirow{4}{*}{MalConvGCG}
& 1\%  & 0.9869 & 0.9871 & $0.1366 \pm 0.0946$ \\
& & 2\%  & 0.9869 & 0.9860 & $0.1212 \pm 0.0747$ \\
& & 5\%  & 0.9869 & 0.9852 & $0.2251 \pm 0.1047$ \\
& & 10\% & 0.9869 & 0.9853 & $0.4557 \pm 0.1870$ \\

\midrule

\multirow{8}{*}{\shortstack[l]{RAMP\\(shift)}}
& \multirow{4}{*}{MalConv}
& 1\%  & 0.9883 & 0.9872 & $0.6538 \pm 0.2063$ \\
& & 2\%  & 0.9883 & 0.9870 & $0.6285 \pm 0.1209$ \\
& & 5\%  & 0.9883 & 0.9862 & $0.7053 \pm 0.1114$ \\
& & 10\% & 0.9883 & 0.9869 & $0.7694 \pm 0.0585$ \\
\cmidrule(lr){2-6}
& \multirow{4}{*}{MalConvGCG}
& 1\%  & 0.9869 & 0.9871 & $0.1392 \pm 0.1135$ \\
& & 2\%  & 0.9869 & 0.9861 & $0.1873 \pm 0.1556$ \\
& & 5\%  & 0.9869 & 0.9866 & $0.1480 \pm 0.0807$ \\
& & 10\% & 0.9869 & 0.9867 & $0.3759 \pm 0.2081$ \\

\bottomrule
\end{tabular*}
\end{table}

\FloatBarrier

\begin{figure}[H]
    \centering
    \begin{subfigure}{0.43\textwidth}
        \centering
        \includegraphics[width=0.95\textwidth]{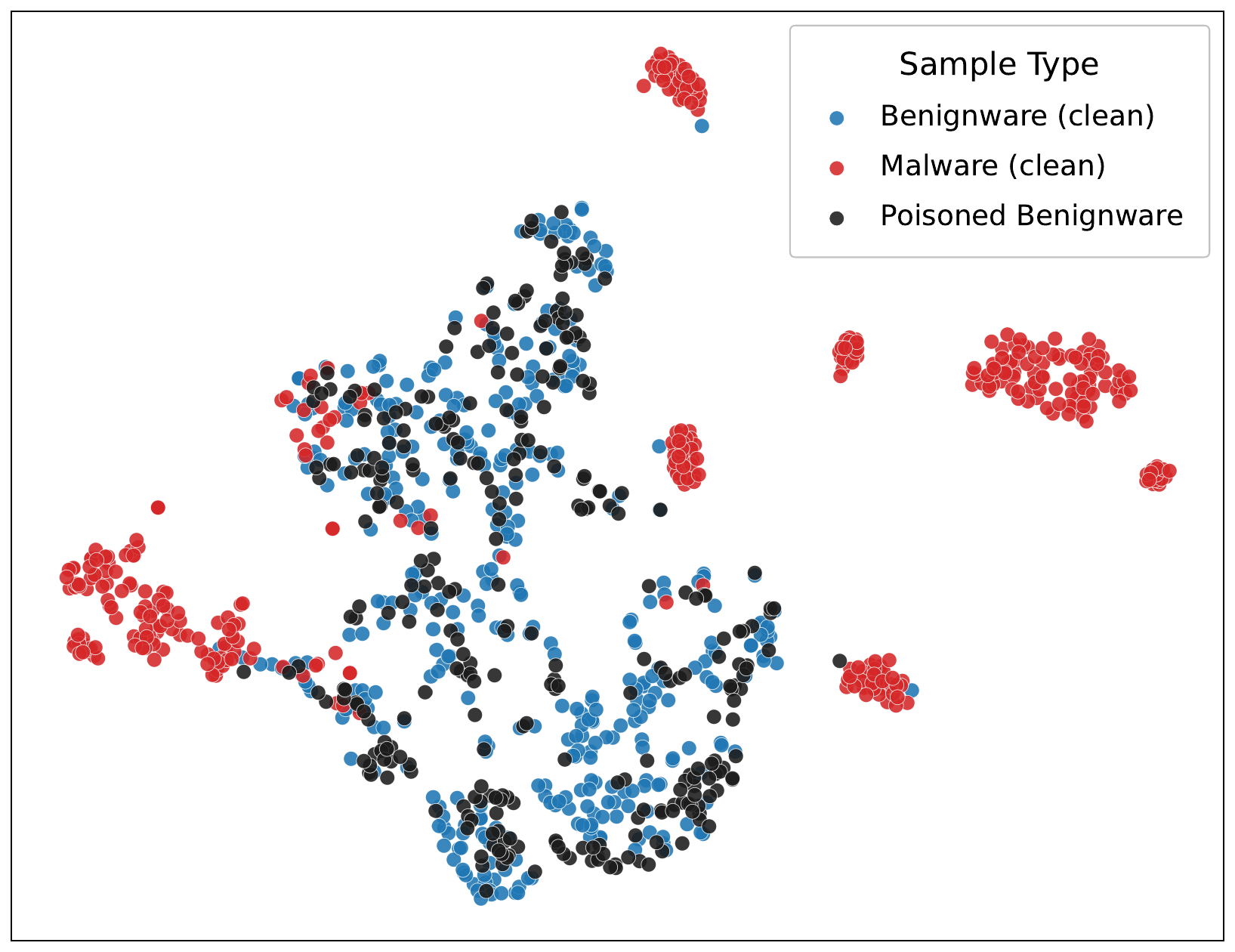}
        \caption{Trigger-only}
        \label{fig:05a}
    \end{subfigure}
    \hfill
    \begin{subfigure}{0.43\textwidth}
        \centering
        \includegraphics[width=0.95\textwidth]{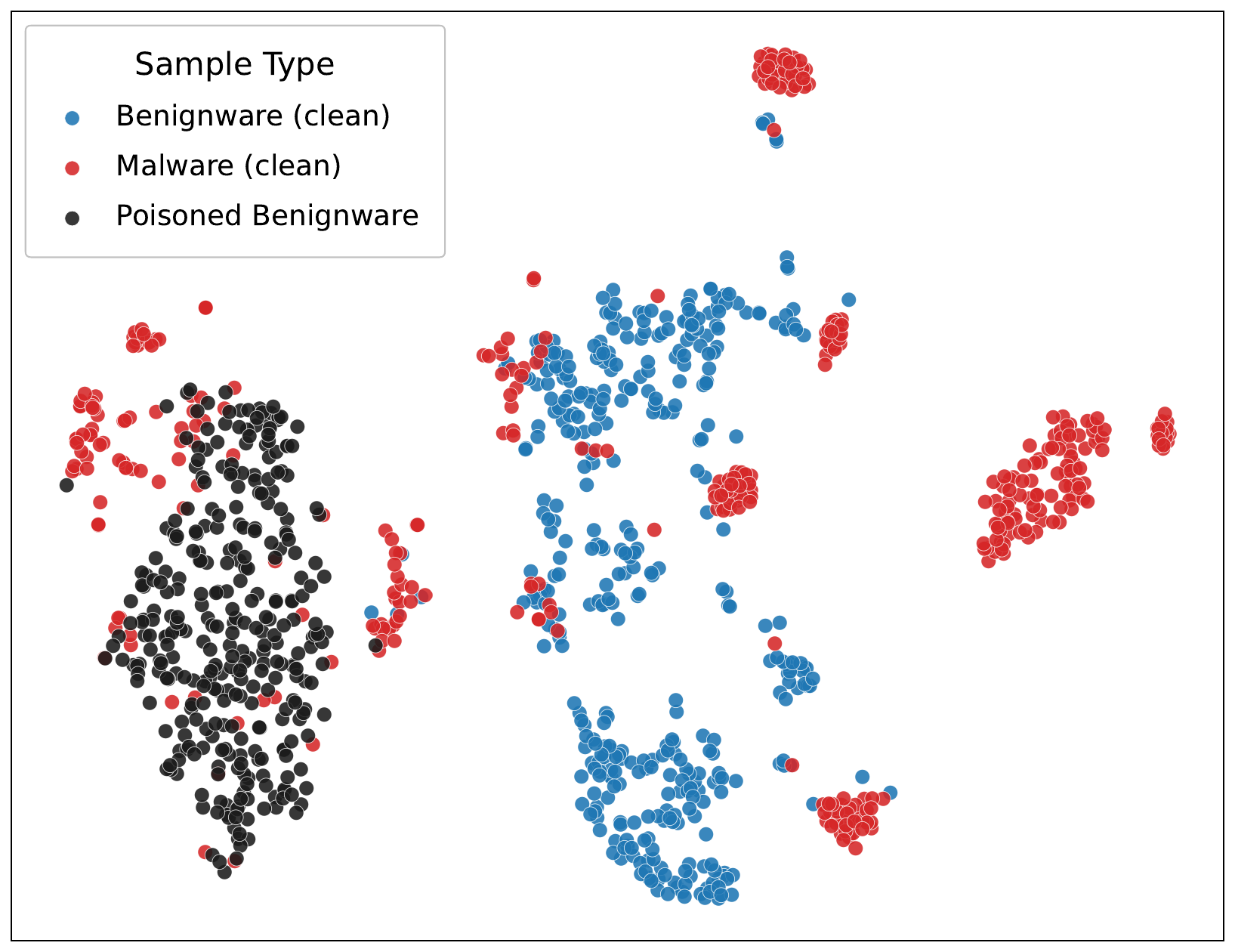}
        \caption{RAMP}
        \label{fig:05b}
    \end{subfigure}
    
    \caption{t-SNE visualization of the feature distributions of trigger-only and RAMP poisoned samples in the pretrained MalConv model.}
    \label{fig:tsne_comparison}
\end{figure}

\section{More Defense Results}

\begin{figure}[H]
    \centering
    \begin{subfigure}[t]{0.48\linewidth}
        \centering
        \includegraphics[width=\linewidth]{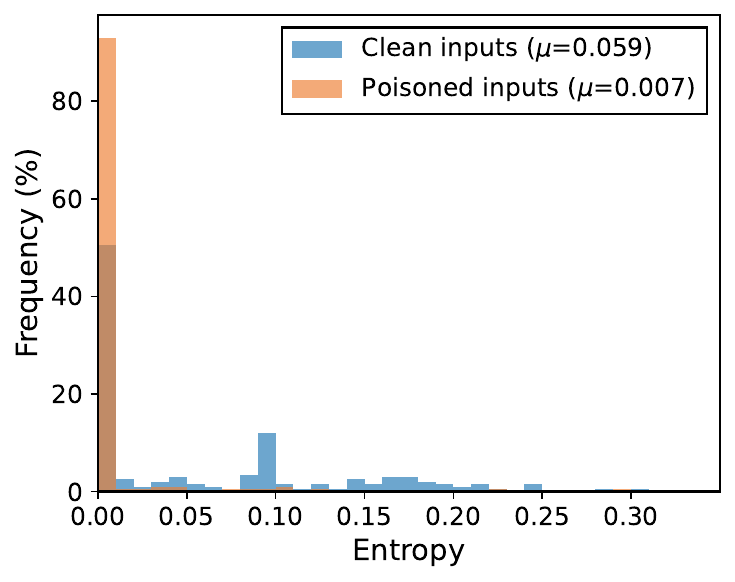}
        \caption{MalConv}
        \label{fig:strip_defense_a}
    \end{subfigure}
    \hfill
    \begin{subfigure}[t]{0.48\linewidth}
        \centering
        \includegraphics[width=\linewidth]{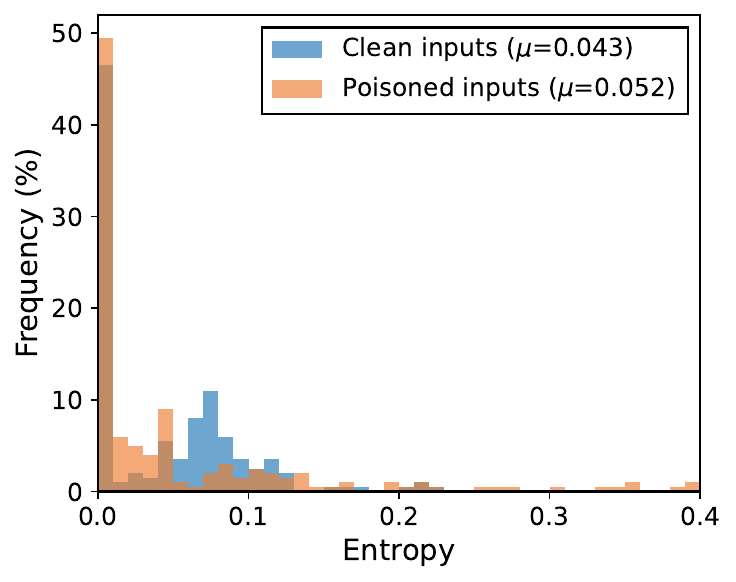}
        \caption{MalConvGCG}
        \label{fig:strip_defense_b}
    \end{subfigure}
    \caption{STRIP entropy distributions on clean and poisoned inputs under RAMP. For each test sample, we generate $N{=}20$ perturbed variants by appending benign overlays and compute the prediction entropy. Each group contains 200 test samples, and $\mu$ denotes the mean entropy.}
    \label{fig:strip_defense}
\end{figure}

\begin{table}[H]
\centering
\scriptsize
\caption{Fine-Pruning results against RAMP. The defense prunes 30\% of low-activation channels in the convolutional layer and then fine-tunes the model on clean data for 5 epochs with a learning rate of $5\times10^{-5}$. ASR Drop is reported in percentage points.}
\label{tab:fine_pruning_defense}
\begin{tabular*}{\columnwidth}{@{\extracolsep{\fill}}l c c c c c c@{}}
\toprule
\multirow{2}{*}{\textbf{Model}} & \multirow{2}{*}{\textbf{PR}} & \multicolumn{2}{c}{\textbf{Before Defense}} & \multicolumn{2}{c}{\textbf{After Fine-Pruning}} & \multirow{2}{*}{\textbf{ASR Drop}} \\
\cmidrule(lr){3-4} \cmidrule(lr){5-6}
 &  & \textbf{CDA} & \textbf{ASR} & \textbf{CDA} & \textbf{ASR} & \textbf{(p.p.)} \\
\midrule
MalConv    & 1\% & 0.9886 & 0.9543 & 0.9848 & 0.8860 & 6.83 \\
MalConvGCG & 1\% & 0.9901 & 0.2193 & 0.9902 & 0.1492 & 7.01 \\
\bottomrule
\end{tabular*}
\end{table}

\begin{table}[H]
\centering
\scriptsize
\caption{ABL results against RAMP under the clean-label DOS setting with a random 96-byte trigger and a poison rate of 1\%. ABL uses an isolation ratio of 1\%, $\gamma=0.5$, 20/10/5 epochs for isolation, fine-tuning, and unlearning, and learning rates of $10^{-3}$, $10^{-4}$, and $5\times10^{-4}$, respectively.}
\label{tab:abl_defense}
\begin{tabular*}{\columnwidth}{@{\extracolsep{\fill}}l c c c c c c@{}}
\toprule
\multirow{2}{*}{\textbf{Model}} & \multirow{2}{*}{\textbf{PR}} & \multirow{2}{*}{\textbf{Isolated / Total}} & \multicolumn{2}{c}{\textbf{Before Defense}} & \multicolumn{2}{c}{\textbf{After ABL}} \\
\cmidrule(lr){4-5} \cmidrule(lr){6-7}
 &  &  & \textbf{CDA} & \textbf{ASR} & \textbf{CDA} & \textbf{ASR} \\
\midrule
MalConv    & 1\% & 0/320 & 0.9886 & 0.9543 & 0.5595 & 0.9985 \\
MalConvGCG & 1\% & 8/320 & 0.9900 & 0.2193 & 0.5000 & 0.0000 \\
\bottomrule
\end{tabular*}
\end{table}

%
% ---- Bibliography ----
%
% BibTeX users should specify bibliography style 'splncs04'.
% References will then be sorted and formatted in the correct style.
%
\bibliographystyle{splncs04}
\bibliography{mybibliography}
%
% \begin{thebibliography}{8}
% \bibitem{ref_article1}
% Author, F.: Article title. Journal \textbf{2}(5), 99--110 (2016)

% \bibitem{ref_lncs1}
% Author, F., Author, S.: Title of a proceedings paper. In: Editor,
% F., Editor, S. (eds.) CONFERENCE 2016, LNCS, vol. 9999, pp. 1--13.
% Springer, Heidelberg (2016). \doi{10.10007/1234567890}

% \bibitem{ref_book1}
% Author, F., Author, S., Author, T.: Book title. 2nd edn. Publisher,
% Location (1999)

% \bibitem{ref_proc1}
% Author, A.-B.: Contribution title. In: 9th International Proceedings
% on Proceedings, pp. 1--2. Publisher, Location (2010)

% \bibitem{ref_url1}
% LNCS Homepage, \url{http://www.springer.com/lncs}, last accessed 2023/10/25
% \end{thebibliography}
\end{document}